\documentclass[a4paper,noshowpacs,showkeys,preprint,superscriptaddress,aps,prb,nobibnotes]{revtex4-2}
\usepackage{graphicx}
\usepackage{amssymb}
\usepackage{amsmath}
\usepackage{epsfig}
\usepackage{natbib}
\usepackage[usenames]{color}

\usepackage[top=1.5in, bottom=1.25in, left=1in, right=1in]{geometry}

\begin{document}

\title{Band-gap tuning in 2D spatiotemporal phononic crystals}

\author{D.~Psiachos}
\email{dpsiachos@gmail.com}
\author{M.~M.~Sigalas}
\email{sigalas@upatras.gr}
\affiliation{Department of Materials Science, University of Patras, 26504, Rio, Greece}

\begin{abstract}
We investigate the effect of small spatiotemporal modulations in subwavelength-dimensioned phononic
crystals with large band gaps, on the frequency
spectrum for elastic waves polarized in the plane of periodicity. When the
radius of cylinders periodically placed inside a matrix of highly-contrasting elastic properties
	is time-varying,
	we find that due to 
	the appearance of frequency harmonics throughout the spectrum, the notion of a
band gap is destroyed in general, although with the appropriate tuning of parameters, in particular the modulation
frequency, it is possible that some band-gap region is retained, making such systems possible candidates for
tunable bandpass filters or phononic isolators, accordingly, and for sensor applications.
\end{abstract}

\maketitle

\section{Introduction}

Phononic lattices
are necessarily engineered structures made up of composite materials (metamaterials). The 
tunability of their design can permit novel behaviour not found in conventional materials,
such as wide, complete band gaps and nonreciprocal propagation~\cite{SigalasReview,CummerReview}.
In the present study, we investigate the effect of spatiotemporal modulation of 
material properties on the band-gap properties for a phononic-lattice
prototype: cylinders embedded in a matrix of significantly
different elastic properties in the subwavelength regime. This type of 
structure is well-known for yielding the large spectral
gaps~\cite{tanaka,sigalas1997,sigalas2000,SigalasReview, kafesaki,Sainidou, Vasseur} desirable in phononic lattices, 
enabling the transmission of one type of polarization or none at all. Recent work 
on two-dimensional colloidal phononic crystals has shown that while 
altering the periodicity of the system results in Bragg gap tuning, 
the appearance as well as the characteristics of band gaps 
above and below the Bragg gap is controllable by altering the
mechanical eigenmodes of the nanoparticles and nanoparticle-membrane
adhesion respectively~\cite{NanoLett}.

Reciprocal propagation of sound or elastic waves in media~\cite{Maznev}, is a general principle 
which assures the interchangeability of 
source and observer: that waves originating from one of these points
propagate in exactly the same manner as waves originating from the other point. It 
is only broken in special situations such as when nonlinearity is present,
when time-reversal symmetry is broken such as by the inclusion of
materials with gain or loss, chirality, angular-momentum bias, 
or in moving systems~\cite{paper15}. It may also be broken when the system itself
is not moving in time but the material parameters are varying in space and time,
manifesting itself as asymmetric band gaps leading to directional 
propagation~\cite{paper4,paper1,paper7,paper18}.

The rectification of the mechanical 
properties of phononic materials has been demonstrated in several experimental
or numerical studies, based on phenomena such as electromagnetism~\cite{wang2018},
piezoelectricity~\cite{paper14,casadei2012,chen2014,ChristensenPRApp}, or
magnetorheological polymers subjected to magnetic fields~\cite{Danas}. For
nonreciprocal propagation to be observed, apart from time-varying material properties, there
must also be spatial variation present.

Some purely mechanical implementations of spatiotemporally-varying material properties 
include a time-varying effective acoustic capacitance achieved in simulations 
involving air-filled waveguides by varying the heights of the attached 
resonators~\cite{paper12,paper10}, a metabeam comprised of multiple resonators with different
reltaive orientations where the resulting stiffness changes in response to 
rotation~\cite{paper11}. Frequency conversion without nonlinearity present can occur
in media with spatiotemporally-modulated
material properties. Depending on the direction of the incident wave in a beam with reflective ends
or a beam containing
an interface between a homogeneous material and one with time-space varying properties, the frequency
can be either up or down-converted, a clear indicator of nonreciprocal 
propagation~\cite{paper8}. The production of harmonics in time-space
modulated media~\cite{paper8,paper13,paper17,paper18} may be exploited for uses
such as unidirectional acoustic isolation~\cite{paper8}, or parametric 
amplification as a possible implementation of a gain medium~\cite{paper13}. In 
an air-filled waveguide system with a vibrating membrane, 
time-varying the membrane tension leads to phenomena such as frequency
conversion; adding spatial variation 
via the addition of a second membrane with a phase 
difference, is what leads to non-reciprocal propagation~\cite{paper9}. Alternatively,
it has been shown that spatial variation leads to the formation of band gaps while the addition of time-variation
is what causes the band structure to display a directionality~\cite{paper5,paper6} via a ``tilting"
of $\omega$-$k$ space, a phenomenon long-known for the case of electromagnetic waves
propagating on a two-dimensional surface with spatiotemporally-modulated properties~\cite{paper2,paper3}.

Given that phononic materials are a very technologically-promising area of research, our 
aim in this study is to examine whether time variation of the phononic crystals' already
spatially-varying material
properties improves or degrades the 
performance of these systems for various objectives such as acoustic isolators or frequency sensors.

\section{Methods}
\begin{figure}[htb]
\includegraphics[scale=0.3]{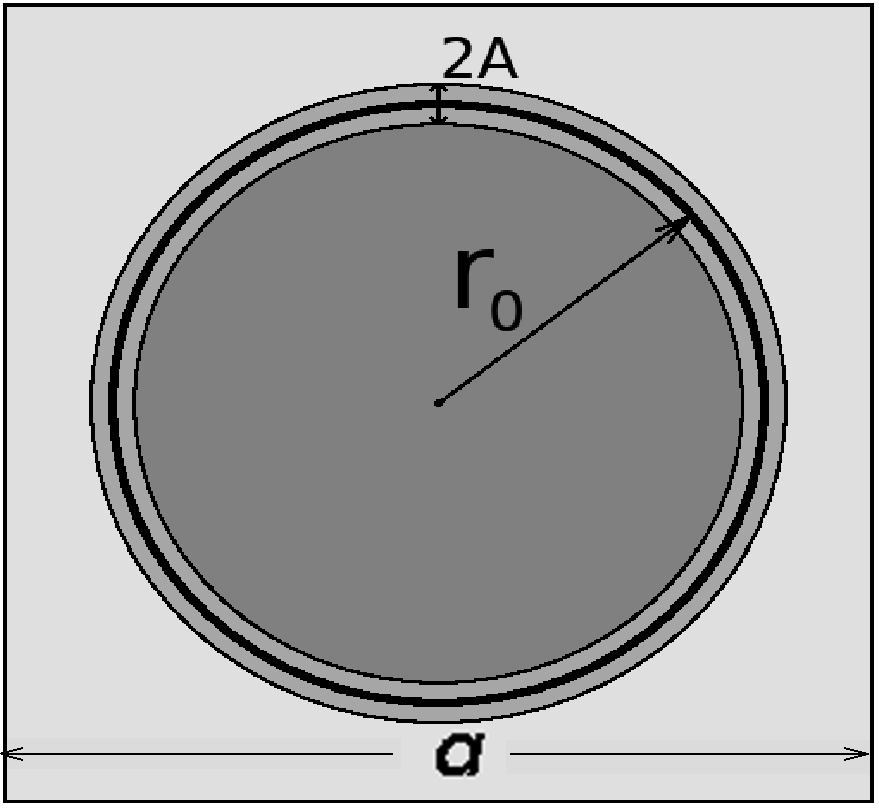}
\caption{Square unit cell in two dimensions with lattice parameter $a$ for the system under study: a composite comprised of a solid cylinder of
	radius $r_0$ inside a background of a different material. The 
	radius of the cylinder varies periodically in time, with a maximum amplitude of $A,$ as depicted.}
\label{model}
\end{figure}

We consider a composite composed of two homogeneous 
structures: a solid cylinder inside a background matrix of a different material, 
together comprising a square unit cell as shown in 
Fig.~\ref{model}. Time-dependent 
material properties in this model are imposed implicitly, via a variation of the radius 
$r=r(t)=r_0+A(t)$ of the cylinder, resulting, due to the nature of the model chosen, in a modification 
of its density as well as the density of the background matrix, as explained further below. The radius
of the cylinder $r_0$ corresponds to a filling factor on the order of 50\%. Such large filling factors in
periodic structures ideally lead to large band gaps~\cite{Sainidou}. The 
function $A(t)=A\sin{(2\pi f t)}$ is a slowly-varying sinusoid with
small modulation amplitude $A$ (up to 2\% of $r_0$), and modulation frequency $f$. This 
constitutes a simplified model of a more realistic implementation of
a system time-varying material properties, such as, for example, a split-ring
structure where the two semicircular components have a different radius, where one is mechanically 
modulated in time inside the other.

\subsection{Formalism}
The elastic wave equation in 
the resulting inhomogeneous elastic medium is
derived from the following relations~\cite{LandauLifshitz}:

a) Mass conservation,
In our model, the cylinder expands and contracts according to
a sinusoidal function, but no mass is actually being added or removed
from the system. Thus, the mass conservation relation
\begin{equation}
        \frac{\partial \rho}{\partial t}+\nabla \cdot (\rho \mathbf{v})=0
        \label{massconserv}
\end{equation}
where $\mathbf{v}= \partial \mathbf{u}/\partial t$ is the velocity in terms
of the displacement vector $\mathbf{u}$, and $\rho$ is the density, holds
for all points in space.

b) Euler's Equation,
\begin{equation}
	\frac{\rho\, \partial v_i}{\partial t}+\rho \sum_j v_j\frac{\partial}{\partial x_j} v_i=\sum_j \frac{\partial T_{ij}}{\partial x_j} 
	\label{momentum}
\end{equation}
where the $i,$ $j$ are the components in space, specifically $x,$$y,$$z,$ of the various 
quantities, $x_i$ are the spatial coordinates, $v_i$ are the components of the velocity, \textit{i.e.} 
$v_i= \partial u_i/\partial t$, for displacements $u_i,$ $T_{ij}$ are the components
of the stress tensor (defined below).

Equations ~\ref{massconserv} and ~\ref{momentum} in conjunction yield
\begin{equation}
       \frac{\partial \left(\rho v_i\right)}{\partial t}=\sum_j \frac{\partial T_{ij}}{\partial x_j}
       \label{conserv}
\end{equation}

after ignoring a term second-order in $\mathbf{v}$ since we aim to work in the regime of small displacements.

c) Hooke's law,
\begin{equation}
	T_{ij}=\lambda u_{ij}\delta_{ij}+2\mu u_{ij}
	\label{hookes}
\end{equation}
where $\delta_{ij}$ is the Kronecker delta function and
\begin{equation*}
	u_{ij}=\frac{1}{2}\left(\frac{\partial u_i}{\partial x_j}+\frac{\partial u_j}{\partial x_i}\right)
\end{equation*}
are the components of the strain tensor $\overline{\overline{u}}$ in terms of the displacement
vector $\mathbf{u}$ and the spatial coordinates $\mathbf{x},$ 
and $\lambda=\rho\left(C_l^2-2C_s^2\right)$ and $\mu=\rho\, C_s^2$ are the Lam\'e coefficients in
terms of the longitudinal and shear sound velocities $C_l$ and $C_s$ respectively, and where 
$\mathbf{u}$ as well as the variables $\rho,$ $\lambda,$ $\mu$ have 
implicit spatial and time dependence governed by the material parameters as determined by
$r(t)$.

\subsection{Finite-Element Time-Domain (FDTD) Calculations}
\label{fdtd}
After an initial disturbance was set in motion at a point 
inside the cylinder, the system was allowed to evolve. The magnitude of the disturbance
is what determines the amplitude of the response and the resulting displacements 
may be considered to remain within
the linear-elastic regime as the problem is completely scalable in this sense.
Using
the finite-difference time-domain (FDTD) method, where the spatial as well as the time domain is discretized, 
and applying periodic boundary conditions and Bloch's theorem~\cite{sigalas2000},
we solve Eqs.~\ref{conserv}-\ref{hookes} for the propagation of elastic waves in 
inhomogeneous media. In particular, the
calculational approach is implemented by discretizing space on a square grid, where the displacements 
$u_x,$ $u_y,$$u_z$ are defined in the centre of each grid cell and their derivatives are approximated
by central-difference formulas in both space and time, resulting in second-order accuracy, while the derivatives for
the material parameters $\rho,$ $\lambda,$ $\mu$ are approximated by 
finite differences between grid and time points. The displacements at the following time step are obtained in terms
of their values, as well as those of other parameters, at previous 
steps, and Fourier-transformed in order to obtain a spectrum
of resonant peaks corresponding to the bands in $k$ space~\cite{sigalas2000}. 

We utilized a square grid of 90 by 90 points for 
the calculations of the band structure for the static model $A(t)=0$ and 180 by 180 points or even denser, up to 
360 by 360,
for more-detailed calculations of the frequency responses of the $k=0$ cases
with and without time dependence because the denser grid let to less abrupt changes in the material parameters 
as a function of time, or led to the 
accessing of smaller amplitudes in $A(t).$ 

All of the lengths are expressed in terms of the lattice parameter $a.$ Frequency 
is expressed as a dimensionless quantity $\tilde{\nu}=\nu a/c$ where $c$ is taken as
the speed of sound $c_b$ in the background material and the units of $\nu$
are \textit{kHz}. The time step in the simulations was taken 
as $\Delta t=$7.698$\times 10^{-4} a/c_b.$

\begin{table}
\begin{tabular}{l|| l| l| l}
	Material& $\rho$ (g/cm$^3$) & $C_l$ (km/s) & $C_s$ (km/s)\\
\hline
	Iron & 7.69 & 5.9 & 3.2\\
	Epoxy & 1.18 & 2.54 & 1.16\\
	Silicon & 2.34 & 8.43 & 5.84
\end{tabular}
	\caption{Densities $\rho$ and sound velocities ($C_l:$ longitudinal and $C_s:$ shear) for
	the materials utilized in the present study.}
\label{tab1}
\end{table}

In Table \ref{tab1} we list the densities and sound velocities of the materials we used for
the composites in our study.

\section{Results}
In this section we apply the formalism and methods outlined in the previous section to two
different composite phononic material systems: iron (Fe) cylinders in epoxy, and silicon (Si) cylinders
in epoxy.
\subsection{Non-time-varying case}
\begin{figure}[htb]
\includegraphics[scale=0.3]{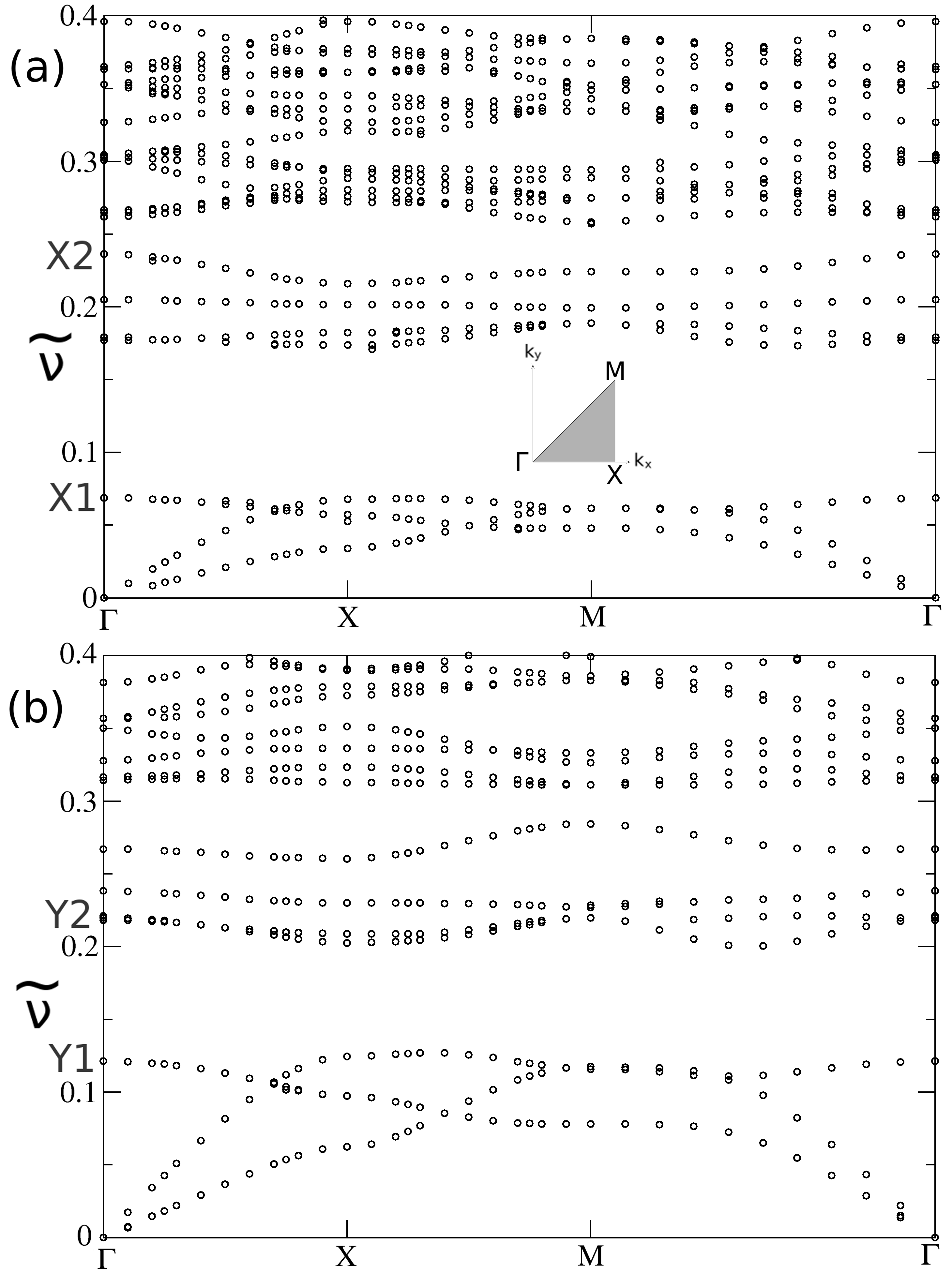}
	\caption{Band structure in two dimensions for elastic waves polarized in the plane of the unit cell 
	(\textit{viz.} Fig.~\ref{model}) for a non-time-varying system with (a) Fe cylinder in an epoxy matrix
	and a filling factor of 50\% and (b) Si cylinder in epoxy for a filling factor of 60\%. 
	Also shown is the irreducible Brillouin zone. The frequency axis is
	dimensionless as explained in the text.}
\label{FeSiEpband}
\end{figure}

The calculated band structure for waves polarized in the plane of the unit cell, of 
mixed transverse-longitudinal character, for 
Fe cylinders in epoxy without time-variation of material properties is shown in 
Fig.~\ref{FeSiEpband}a, for a filling factor of 50\%. The plot is shown for a dimensionless
frequency $\tilde{\nu}$. The large band gaps are what make these types of materials ideal phononic materials. The
ratio of the gap-width defined as $\Delta\omega$ to the mid-gap frequency defined as $\omega_g,$ for 
the largest gap, $\Delta\omega/\omega_g,$ is 0.86
while for a filling factor of 60\% for the same materials, the ratio is 0.91. 

Similarly, for Si cylinders in epoxy without time variation we also find a band 
structure with large gaps, as shown in Fig.~\ref{FeSiEpband}b,
for a filling factor of 60\%. The gap-width to midgap frequency ratio $\Delta\omega/\omega_g$ 
for the largest gap is 0.45, much smaller than for Fe in epoxy even at the same
filling factor, a result explained by the smaller contrast in the densities between the cylinder 
and matrix for this case, a factor
of greater significance in determining this ratio than the disparity
between the elastic constants~\cite{KushwahaApp}, which in our case, is about the same
for both composites. 

\subsection{Time-varying material properties}

\begin{figure}[htb] 
\includegraphics[scale=0.3]{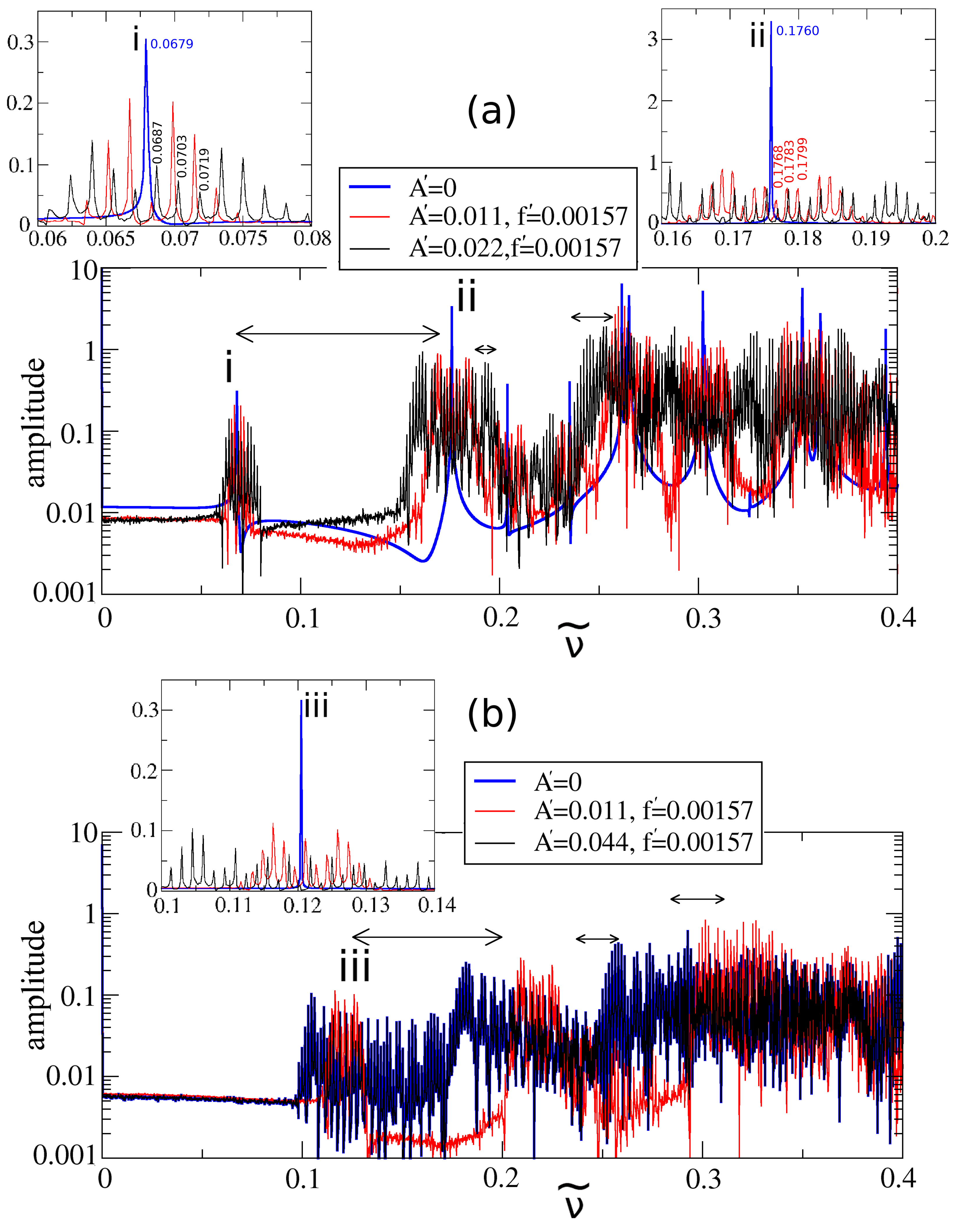}
	\caption{Frequency spectrum for a point-source excitation for a system with, (a) an Fe cylinder 
	in an epoxy matrix and a filling factor of 50\% and (b) a Si cylinder in epoxy
	at a filling factor of 60\%, at $k=0$. Shown is the spectrum without
	any time variation of cylinder radius ($A^\prime=0$)
	as well as with nonzero maximum amplitude modulations $A^\prime$, at a frequency of 
	0.00157 (dimensionless units). The horizontal arrows show the locations of the absolute
	band gaps from the non-time-varying case (Fig.~\ref{FeSiEpband}). The insets show in detail
	the structure of the lowest resonance peaks in each system (i-ii), wherein for Fe
	in epoxy some of the frequencies are labelled,
	as explained in the text, in the colour corresponding to
	the respective modulation parameters.}
\label{FeSiEpampl}
\end{figure}

As we time-vary the cylinder radius as outlined in the previous section, for an Fe or a Si 
cylinder in epoxy, we obtain the spectrum shown in 
Fig.~\ref{FeSiEpampl} for various maximum amplitude variations of a sinusoid and fixed modulation frequency, scaled
respectively to be dimensionless as $A^\prime=A/a$ and $f^\prime=fa/c,$ at
$k=0$. The vertical axis depicts the amplitude of the frequency spectrum  
$\sqrt{u_x(\tilde{\nu})^2+u_y(\tilde{\nu})^2},$ obtained after Fourier-transforming
the value of the displacement at a fixed point inside the cylinder, near its centre. The units
for the displacement are arbitrary as mentioned in Sec.~\ref{fdtd} owing to the scalability
of the problem as far as the amplitude of the initial pulse is concerned. The 
horizontal arrows denote
the location of the absolute band gaps in the non-time-varying case (\textit{viz.} 
Fig.~\ref{FeSiEpband}). Most notable are the blueshifting of the main resonance peaks, as well 
as the appearance of accompanying sidebands upon adding time-modulation, offset from 
the main resonances by the amount of the modulation
frequency, and the existence higher harmonics fading with distance in intensity. The sidebands do not necessarily
have a smaller amplitude than the main resonance peak. As the modulation amplitude is increased, the
resonance peaks become increasingly blueshifted and the sidebands cover a wider
frequency range, as can be seen clearly by the insets for the region around the resonance peaks. In 
the case of Fe in epoxy (Fig.~\ref{FeSiEpampl}a), the two lowest peaks in the non-time-varying case 
corresponding to a frequency of 0.0679 and 0.1760 are shifted towards the values of 0.0687 and
0.1768 respectively for $A^\prime$ being 0.011 and 0.022 respectively, as shown
in the insets, while the respective sidebands are separated from the respective peaks 
by multiples of the value $f^\prime=0.00157,$ as shown, to within an excellent approximation. Further
details, involving smaller amplitude modulations and regarding the amplitudes of the displacements
upon adding time-modulation, calculated for the system of an Fe cylinder in epoxy, for a larger 
filling factor, 60\%, may 
be found in the Appendix.

In Fig.~\ref{FeSiEpampl}b, which depicts the frequency spectrum for an Si cylinder in epoxy
at 60\% filling, the largest $A^\prime$ depicted is two times larger than that shown in 
Fig.~\ref{FeSiEpampl}a for Fe cylinders in epoxy and all of the band gaps are completely obliterated although
it is clear that the harmonics destroy the band gaps in all cases - especially as a result of the contribution
of other wavevectors given that the bands themselves are not flat. However, for small amplitude modulations,
the harmonics do die off quickly and in principle, some region within the largest band gap would be 
left intact. We note that the bands for Fe in epoxy are flatter than for Si in epoxy in the non-time
varying case (\textit{viz.} Fig.~\ref{FeSiEpband}), something which
occurs even for the same filling factors and whose origin lies in the greater hybridization of the 
continuum bands with the localized resonances from rigid-body modes of the cylinder in the case
of Fe, due to the sharper contrast in the densities~\cite{Kushwaha1998,Sainidou}.

\begin{figure}[h]
\includegraphics[scale=0.3]{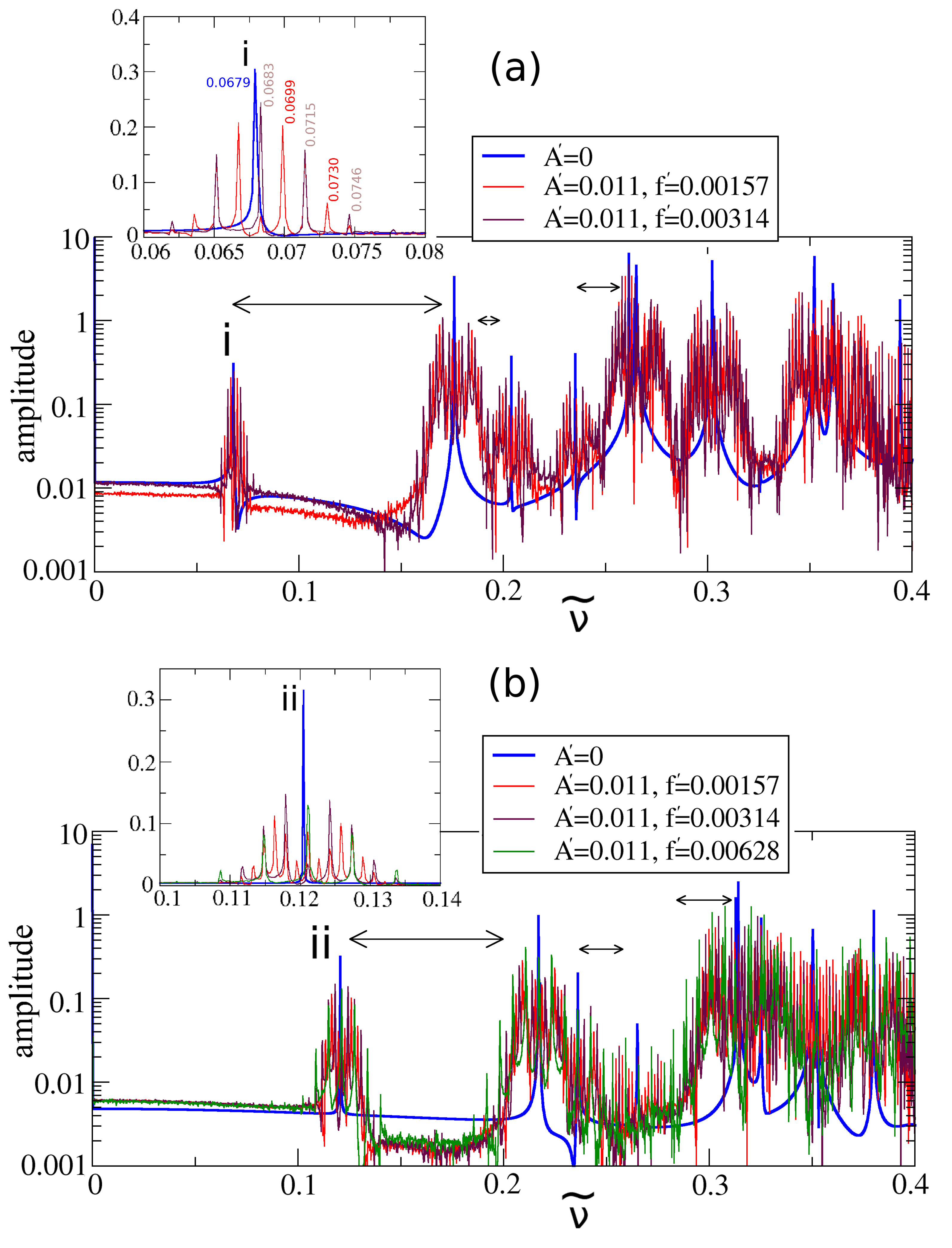}
	\caption{Frequency spectrum for a point-source excitation for a system with, (a) an Fe cylinder 
	in an epoxy matrix and a filling factor of 50\%, (b) a Si cylinder in epoxy
	and a filling factor of 60\%, at $k=0$. Shown is the spectrum without
	any time variation of cylinder radius ($A^\prime=0$)
	as well as with a maximum amplitude modulation of 0.011, for different frequency
	modulations. The horizontal arrows show the locations of 
	the absolute band gaps from the non-time-varying case (Fig.~\ref{FeSiEpband}). The insets show in detail
	the structure of the lowest resonance peak in each system (i-ii) wherein for Fe in epoxy some of the
	frequencies are labelled, as explained in the text, in the colour corresponding to
	the respective modulation parameters, except for the brown labels which also indicate the position of
	the $f^\prime=0.00157$ resonances, which always coincide with the $f^\prime=0.00314$ resonances.}
\label{FeSiEpfreq}
\end{figure}

For fixed amplitude modulation and varying frequency modulation of the cylinder radius, we obtain
the results in Fig.~\ref{FeSiEpfreq}, for the two systems, confirming that the sidebands  
are offset by an amount equal to the modulation frequency $f^\prime$ and that the blueshift of the main
resonance peak compared to the non-time-varying case is unaffected by the value of $f^\prime$. In the
case of Fe in epoxy in Fig.~\ref{FeSiEpfreq}a, the lowest peak in the non-time varying case
corresponding to a frequency of 0.0679 is shifted towards the value of 0.0683 for both
$f^\prime=0.00157$ and 0.00314 for $A^\prime=0.011$ while the sidebands appear at successive separations
of $f^\prime$ from the main peak. The results for the frequency variation confirm that upon adding
time-variation, a blueshift is in fact occurring because, as the resonances for different
frequency modulation but the same amplitude modulation coincide, there should be a resonance
from both modulation frequencies immediately to the left
of the nonmodulated resonance in the inset of Fig.~\ref{FeSiEpampl}a if the resonances
were redshifted instead of blueshifted, 
as this peak would have corresponded to the main resonance as opposed to a sideband.

As it constitutes a more realistic implementation of a time-varying radius, we repeated
some of our calculations on a simplied model of a mechanically-modulated split-ring
structure, that of a hollow cylinder. The results were similar, with
large band gaps for a non-time-varying radius and resonances with harmonics when the outer
radius is periodically modulated in time. The details and results of out calculations are shown in the Appendix.

\subsection{Displacement profiles}
\begin{figure}[htb]
\includegraphics[scale=0.4]{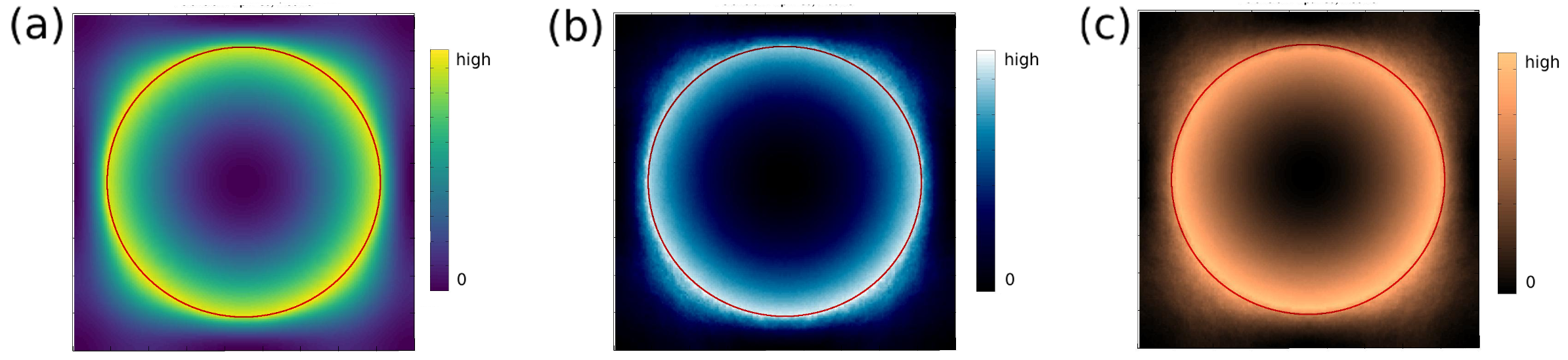}
        \caption{Displacement profile $u_x^2+u_y^2$ projections in the plane of periodicity, at the 
	lowest resonance peak for a Fe cylinder in epoxy at a 50\% 
	filling factor at $k=0$ (see point X1 in Fig.~\ref{FeSiEpband}a and peak (i) Fig.~\ref{FeSiEpampl}a) for (a) 
	a non-time-varying
	cylinder radius, and (b)-(c) for the time-varying cases of $A^\prime=0.022$ and 
	$f^\prime=0.00157$ respectively. The excitation
        was performed by a plane wave with a frequency of 0.0681 for (a) while in (b) and (c) the excitation frequencies
	were 0.0689 and 0.0705 (in dimensionless units $\tilde{\nu}$), corresponding to the main 
	resonance and first sideband, respectively. The 
	colour maps
        are arbitrary - the results depicted are not on the same scale. Shown also is the circle
	outlining the outer boundary of the cylinder.}
\label{localFeEp}
\end{figure}

\begin{figure}[htb]
\includegraphics[scale=0.4]{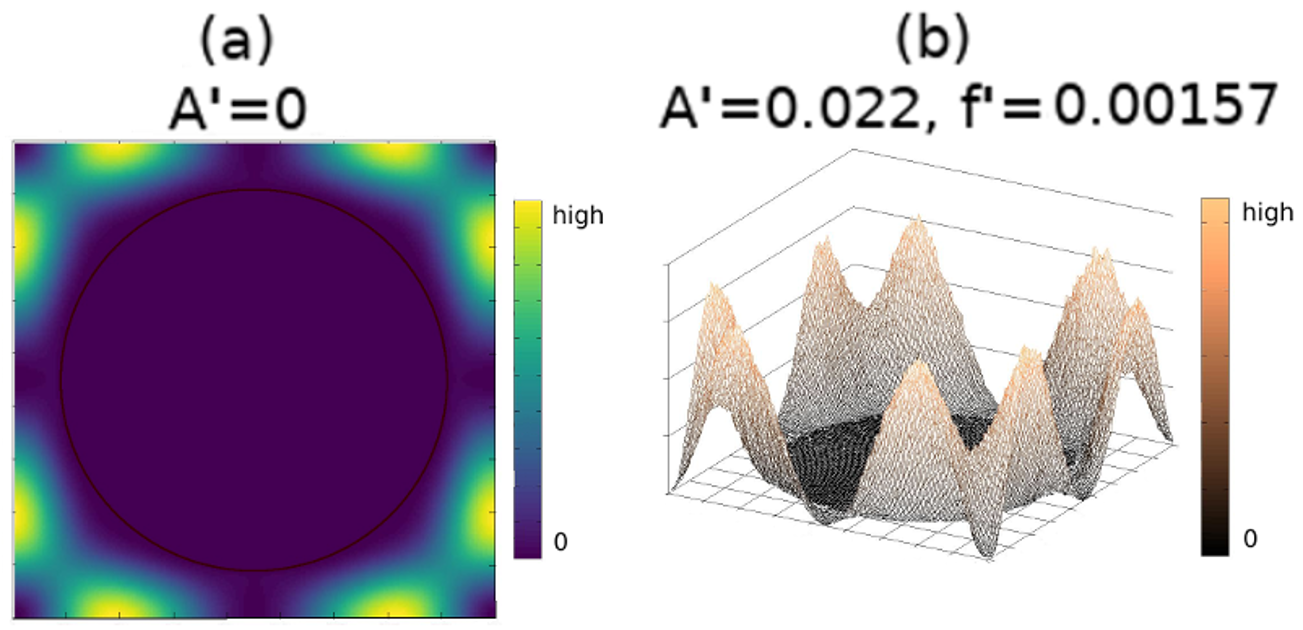}
	\caption{Displacement profile of the point labelled as X2 in Fig.~\ref{FeSiEpband}a ($\tilde{\nu}=0.2352$)
	at $k=0$ for a Fe cylinder in epoxy at a 50\% filling factor, for (a) a non-time-varying
	cylinder radius (plane projection) and (b) time-varying case (profile view) whereupon
	the resonance frequency is shifted to $\tilde{\nu}=0.2361.$ The red circle
	denotes the boundary of the cylinder.} 
	\label{localFeEp597}
\end{figure}

Upon adding time-variation to the material properties, we find that the localization 
character of the vibrations associated with each resonance remains unchanged;
even up to the second harmonic, it remains constant with the only difference from the non-time-varying
case being that the surrounding material is also disturbed. In Fig.~\ref{localFeEp} we show
the displacement profile $u_x^2+u_y^2$ of the lowest resonance peak at the 
frequency $\tilde{\nu}=0.0681$ in (a) the non-time-varying 
case (point X1 in Fig.~\ref{FeSiEpband}a), 
and (b) its counterpart for the case where the maximum amplitude is $A^\prime=0.022$ and frequency 
$f^\prime=0.00157$ where the main
peak is now at a frequency of $\tilde{\nu}=0.0689$. We also show, in Fig.~\ref{localFeEp}c, 
the localization of the first sideband on the 
right having a frequency of $\tilde{\nu}=0.0705$. In all of these cases, the vibrations were highly localized
on the outer edge of the cylinder, albeit blurred in the case of the time-varying case. In all of the 
higher-frequency resonances, the vibrations were localized outside the cylinder. For example,
for point X2 in Fig.~\ref{FeSiEpband}a, at a reduced frequency of $\tilde{\nu}=0.2352$ without time-variation, we have  
Fig.~\ref{localFeEp597}a which is only slightly disturbed in Fig.~\ref{localFeEp597}b, by adding amplitude and frequency 
modulation of $A^\prime=0.022$ and $f^\prime=0.00157$ respectively, whereupon
the resonance frequency shifts slightly to $\tilde{\nu}=0.2361$. In all of the above
situations, the displacement profile was generated by exciting with a plane wave at the frequency being
investigated, and examining the square of the displacement in the plane $u_x^2+u_y^2,$ at steady state once the excitation has long died out. During
the steady state, there are of course vibrations but the localization character of each resonance remains unchanged.

\begin{figure}[htb]
\includegraphics[scale=0.4]{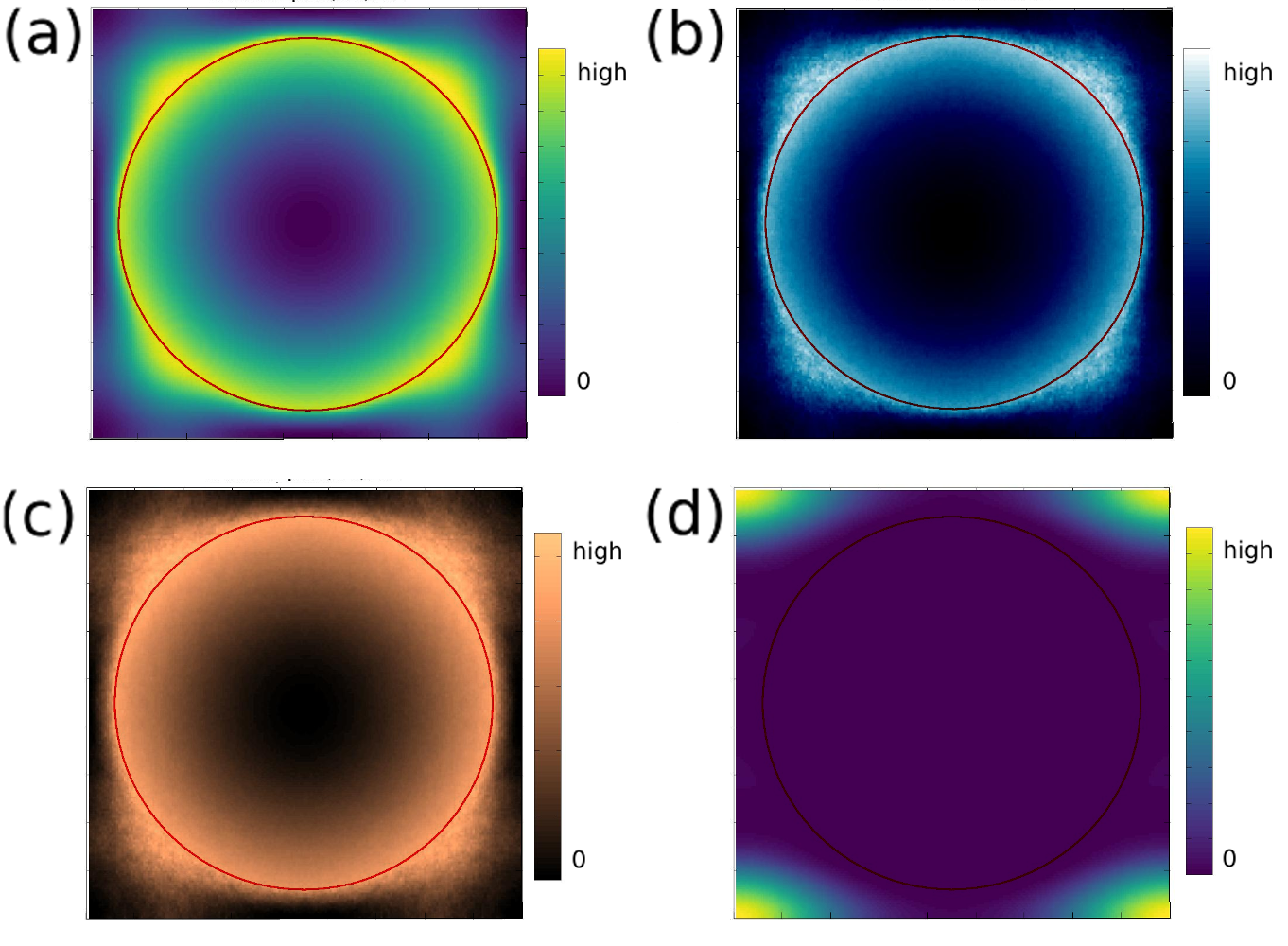}
	\caption{Displacement profile at the lowest resonance peak for a Si cylinder in epoxy
	at a 60\% filling factor at $k=0$ (at the point Y1 in Fig.~\ref{FeSiEpband}b and also 
	peak(iii) in Fig.~\ref{FeSiEpampl}b) for a non-time-varying
	cylinder radius: (a), and for the time-varying case of $A^\prime=0.044$ and 
	$f^\prime=0.00157$: (b)-(c). The excitation was performed by a plane wave with a frequency of 
	$\tilde{\nu}=0.1205$ for (a) while in (b) and (c) the excitation frequency was 0.1217 and 0.1248 
	respectively, corresponding to the main resonance and the second sideband respectively. In (d)
	the plane projection of the vibrational	displacement for the second-lowest resonance for the non-time-varying case 
	(point Y2 in Fig.\ref{FeSiEpband}b occuring at an excitation frequency of $\tilde{\nu}=0.2169$ is
	depicted. The colour maps
	are arbitrary - as the results depicted are not on the same scale. Shown also is the circle
	outlining the outer boundary of the cylinder.}
\label{localSiEp}
\end{figure}

In Fig.~\ref{localSiEp} we plot the displacement profile $u_x^2+u_y^2$ for a Si cylinder in epoxy 
for the lowest resonance peak at the frequency 0.1205 in 
the non-time-varying case (point Y1 in Fig.~\ref{FeSiEpband}b), and in order to compare it 
with its counterpart at the frequency 0.1217 in the time-varying case
where $A^\prime=0.044$ and $f^\prime=0.00157$ (\textit{viz.} inset in Fig.~\ref{FeSiEpfreq}b) 
in Fig.~\ref{localSiEp}b. In Fig.~\ref{localSiEp}c
we show the second harmonic of the lowest resonance at $\tilde{\nu}=0.1248.$ Similar to the case
of Fe in epoxy, the vibrations are 
highly localized on the edge of the Si cylinder (outlined in the figures)
although it is markedly blurry in the time-varying case. There are bulges near the corners of the unit cell
which are associated with the proximity of the neighbouring cylinders. They are not present in the case
of the Fe cylinder in epoxy depicted earlier because there the filling factor was 50\%, but they are seen
in that system as well for a filling factor of 60\%. As for the case of Fe in epoxy, for the higher
resonances, the vibrations are localized in the matrix outside the cylinder. In 
Fig.~\ref{localSiEp}d we show the localization
of the vibrations for the point Y2 in Fig.~\ref{FeSiEpband}b located at $\tilde{\nu}=0.2169.$ Point Y2
actually corresponds to two very nearly degenerate frequencies, with virtually-identical localization 
profiles. Just one of these is shown in Fig.~\ref{localSiEp}d, with more details provided
in the Appendix. Due to the
close proximity of the cylinders at this large filling factor in neighbouring unit cells, the vibrations
of this resonance are preferentially localized around the corners of the unit cell, where the
spacing between neighbouring cylinders is greatest.

\section{Conclusions}
We have examined the effect of periodic strucural modulations in two phononic materials composed
of a cylinder inside a matrix of highly-contrasting elastic properties and we have found
that in general the large band gaps are destroyed by harmonics generated by the temporal
modulation. However, the resonances' time-harmonics 
induced by structural modulations may be exploited
in order to achieve tunable phononic isolation or alternatively, bandpass filtering given that 
in phononic systems with fairly flat bands, by employing a large-enough modulation frequency, it is feasible that
small band gaps could be retained for some regions. The harmonics'
periodicity and wide range could also find implementation in a 
frequency sensor. Modulating the cylinders' radius may be achieved
relatively simply: such as by a mechanically-modulated split-ring structure where the two components have
a different radius.

\section{Appendix}
\makeatletter
\renewcommand{\thefigure}{A\@arabic\c@figure}
\makeatother
\setcounter{figure}{0}    
\subsection{Fe cylinder in Epoxy, filling factor of 60\%}
\begin{figure}[htb]
\includegraphics[scale=0.3]{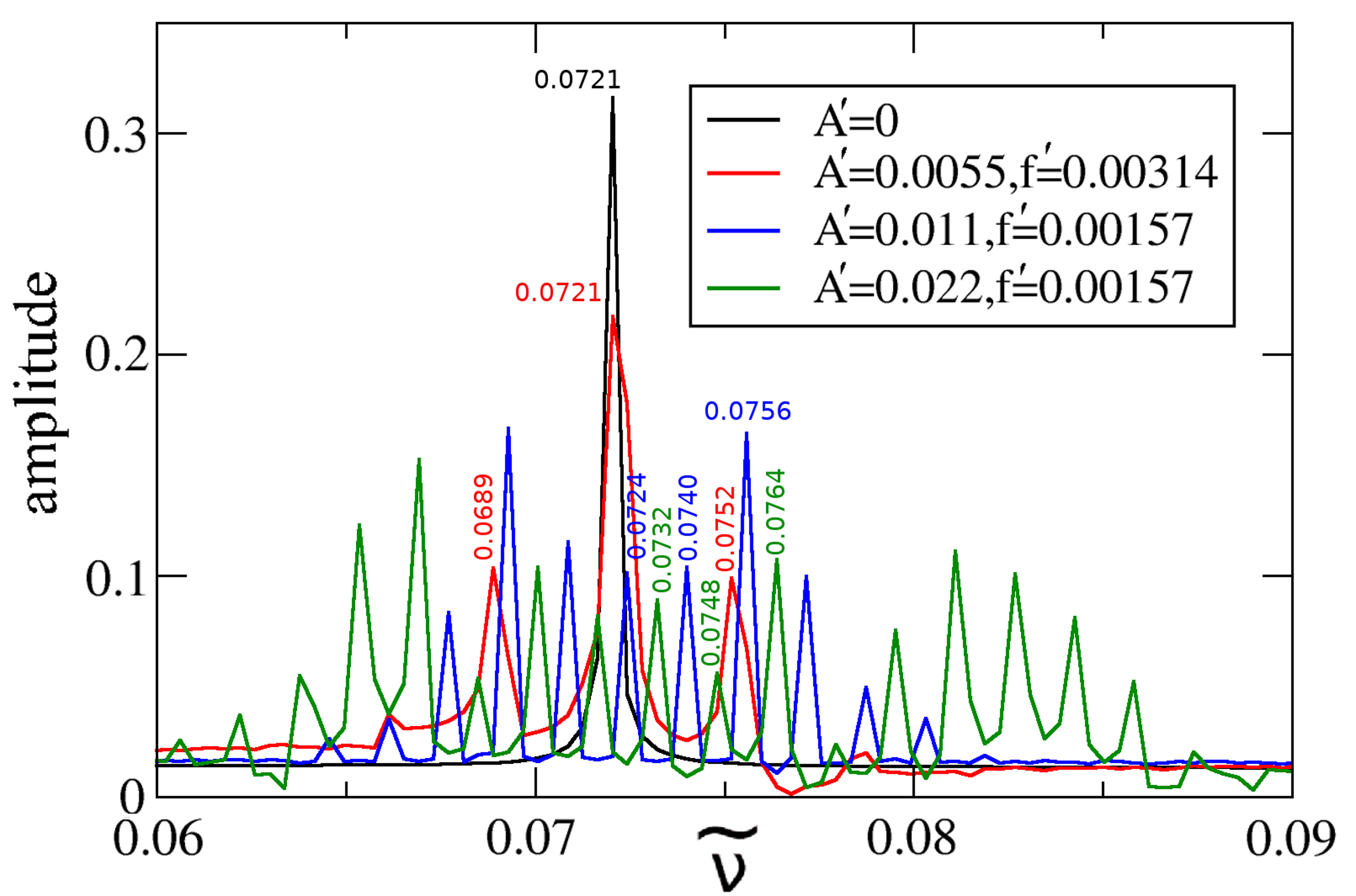}
        \caption{Lowest resonance of the frequency spectrum for a point-source excitation
	for a system with an Fe cylinder
        in an epoxy matrix and a filling factor of 60\%, at $k=0$. Shown
	is the spectrum without
        any time variation of cylinder radius ($A^\prime=0$)
        as well as with nonzero maximum amplitude modulations $A^\prime$, and two
	different frequencies. The dimensionless frequencies for some peaks are
	labelled, in the colour of the corresponding values of ($A^\prime,f^\prime$).}
\label{A1}
\end{figure}

In Fig.~\ref{A1} we show  results for the frequency
spectrum for the lowest resonance at $k=0$ (analogous to point X1 in Fig.~\ref{FeSiEpband}a in the
main text) for the system of an Fe cylinder in epoxy for a filling factor of 60\% for a
time-varying cylinder radius for different maximum amplitude and frequency variations.
The blueshift for the smallest amplitude variation ($A^\prime$=0.0055) is not discernible
for the calculational parameters used but it is evident for the larger $A^\prime.$ In
general, the results are similar to the case of 50\% filling described in the main text. In
the case of a time-varying cylinder radius,
sidebands and higher harmonics appear around the main resonance peak, separated by the value
of the dimensionless frequency $f^\prime$.

\begin{figure}[htb]
\includegraphics[scale=0.3]{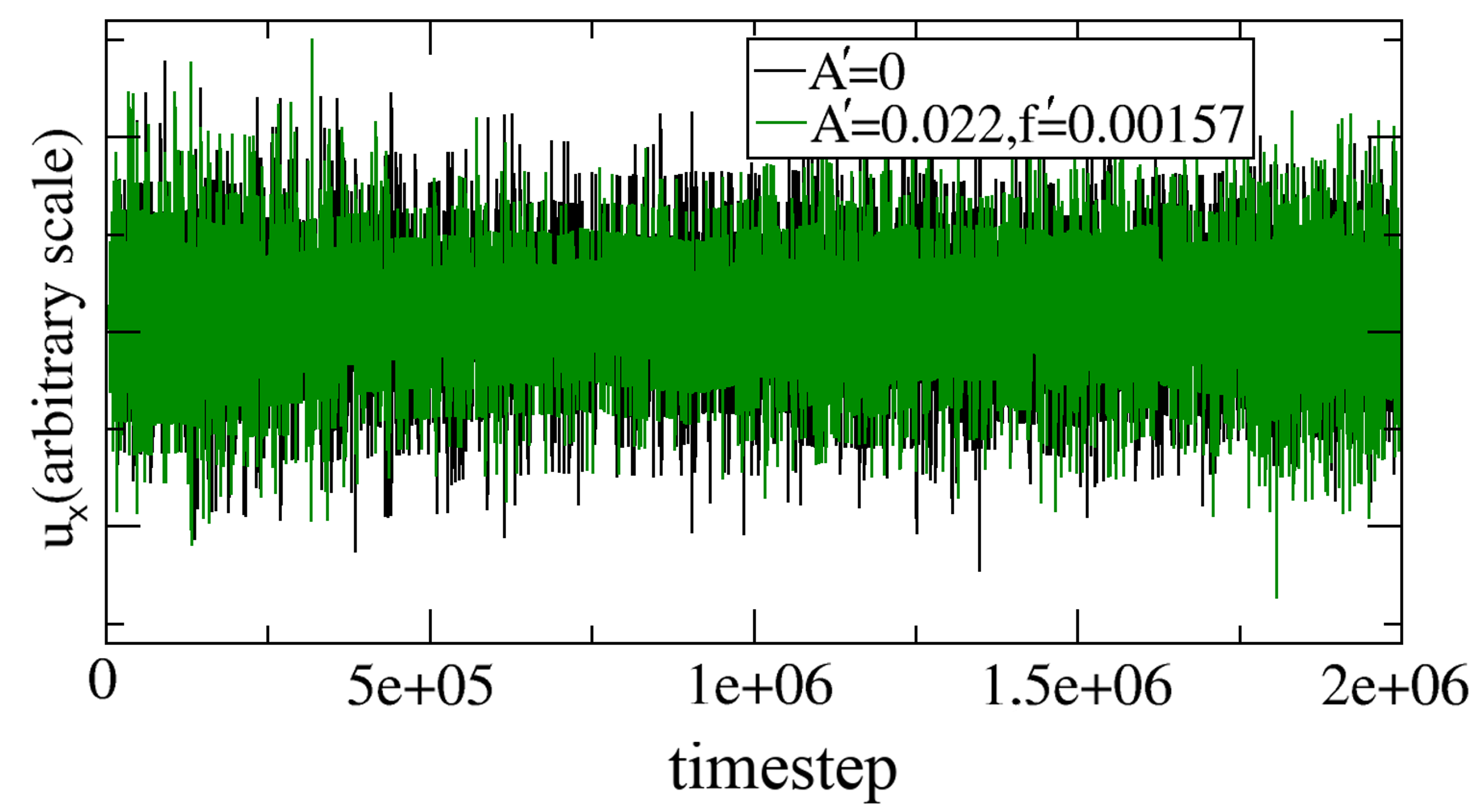}
	\caption{Displacement $u_x$ (arbitrary scale) as a function of timestep after initial excitation
	of the system of an Fe cylinder in epoxy and a filling factor of 60\% for a fixed ($A^\prime$=0)
	and time-varying cylinder radius.}
\label{A2}
\end{figure}

In Fig.~\ref{A2} we show, for the same system, for a fixed as well as for a
time-varying cylinder radius, the variation of the material displacement at a
point inside the cylinder, as a function of time, where the timestep is
$\Delta t=$7.698$\times 10^{-4} a/c_b$ for $a$ being the lattice parameter and $c_b$
the speed of a longitudinal sound wave in the background medium (epoxy). The magnitude
of the displacement is entirely scalable as it varies linearly with the magnitude
of the initial pulse used to excite a point inside the system. Therefore, the displacements
may in this manner be constrained to the regime of linear elasticity. As can be seen from the
figure, the displacements are on the same scale either for a fixed or for a time-varying cylinder
radius.

\subsection{Hollow Cylinder of Fe in Epoxy}

\begin{figure}[htb]
\includegraphics[scale=0.3]{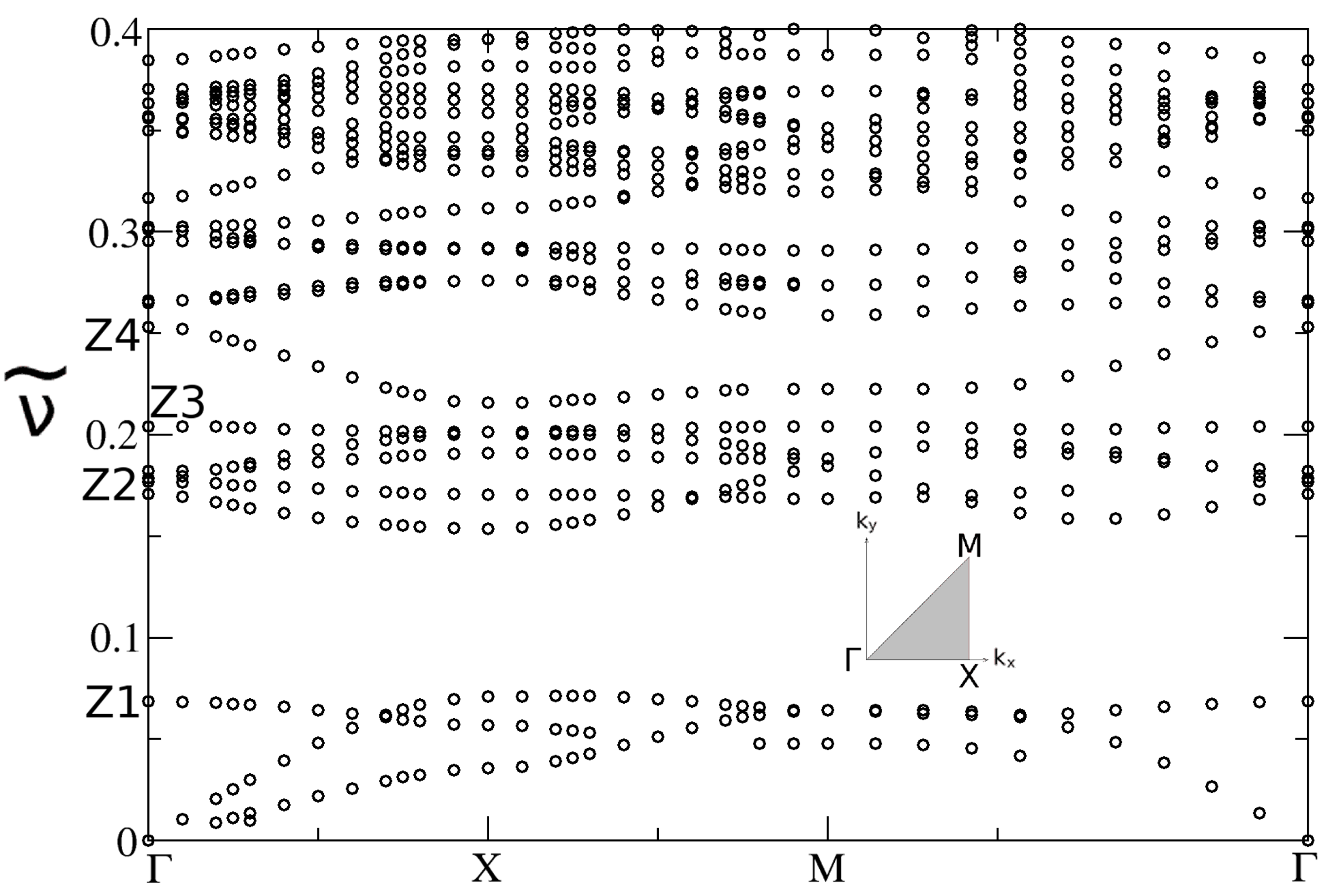}
        \caption{Band structure in two dimensions for elastic waves polarized in the plane of the unit cell
        for a non-time-varying system with an Fe hollow cylinder in an epoxy matrix. The outline
	of the cylinder is as shown in Fig.~\ref{A5}. Also shown is the irreducible
	Brillouin zone. The frequency axis is dimensionless as explained in the text.}
\label{A3}
\end{figure}

\begin{figure}[htb]
\includegraphics[scale=0.4]{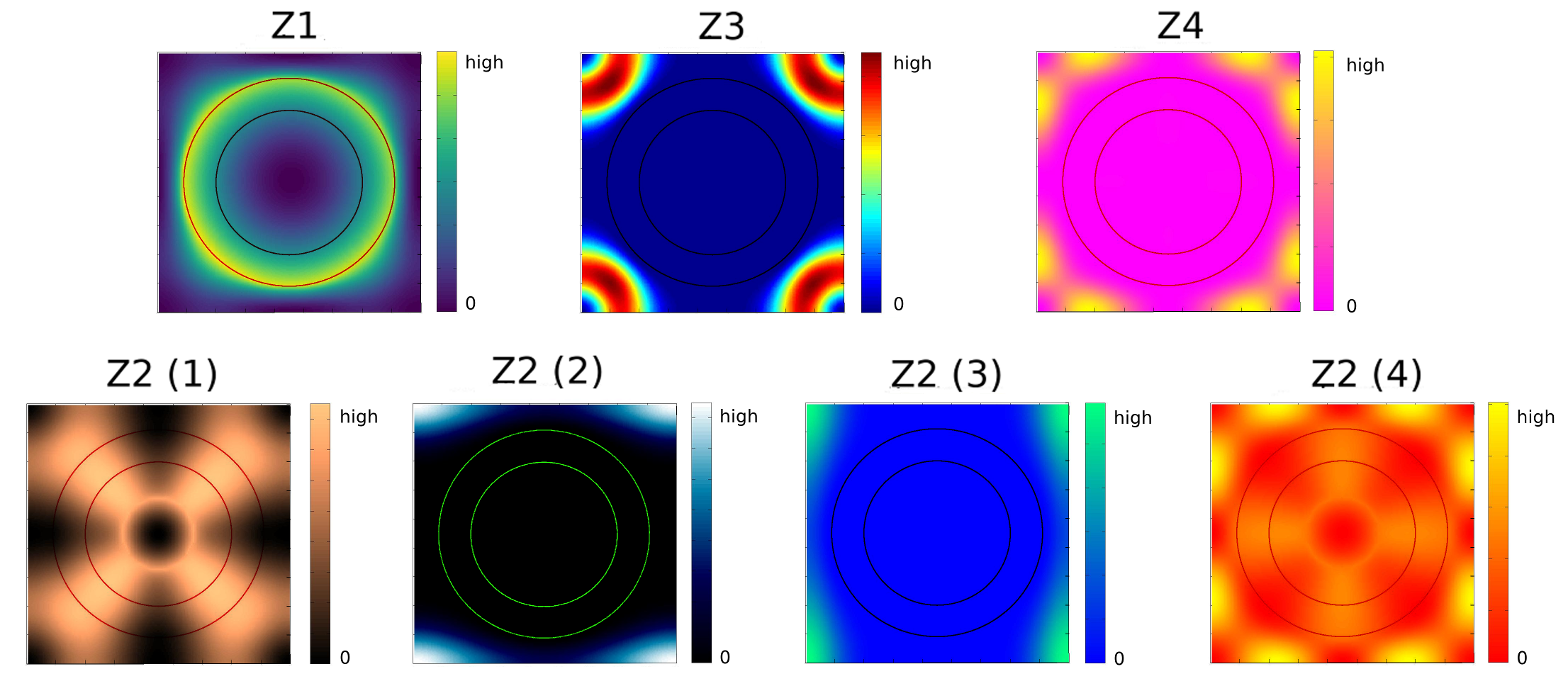}
	\caption{Displacement profiles (plane projection) of the points labelled as Z1-Z4 in Fig.~\ref{A3}
	at $k=0$ for an Fe hollow cylinder in epoxy. The point Z2 in Fig.~\ref{A3} corresponds to four
	closely-spaced frequencies. Superimposed is the outline of the ring. All images are for a
	non-time-varying cylinder radius. The maximum amplitudes of the colour scales are
	arbitrary.}
        \label{A5}
\end{figure}

If, instead of a solid cylinder as in the main text, we have
a hollow cylinder of Fe in Epoxy with the same outer
dimension as the cylinder described in the main text (of 50\% filling factor) and
inner dimension as shown by the outlines in Fig.~\ref{A5}, we obtain the band structure
of Fig.~\ref{A3} for elastic waves polarized in the plane of the unit cell for a non-time-varying
system. In general, the band structure is very similar to its counterpart for a solid cylinder
(Fig.~\ref{FeSiEpband}a in the main text).

\begin{figure}[htb]
\includegraphics[scale=0.3]{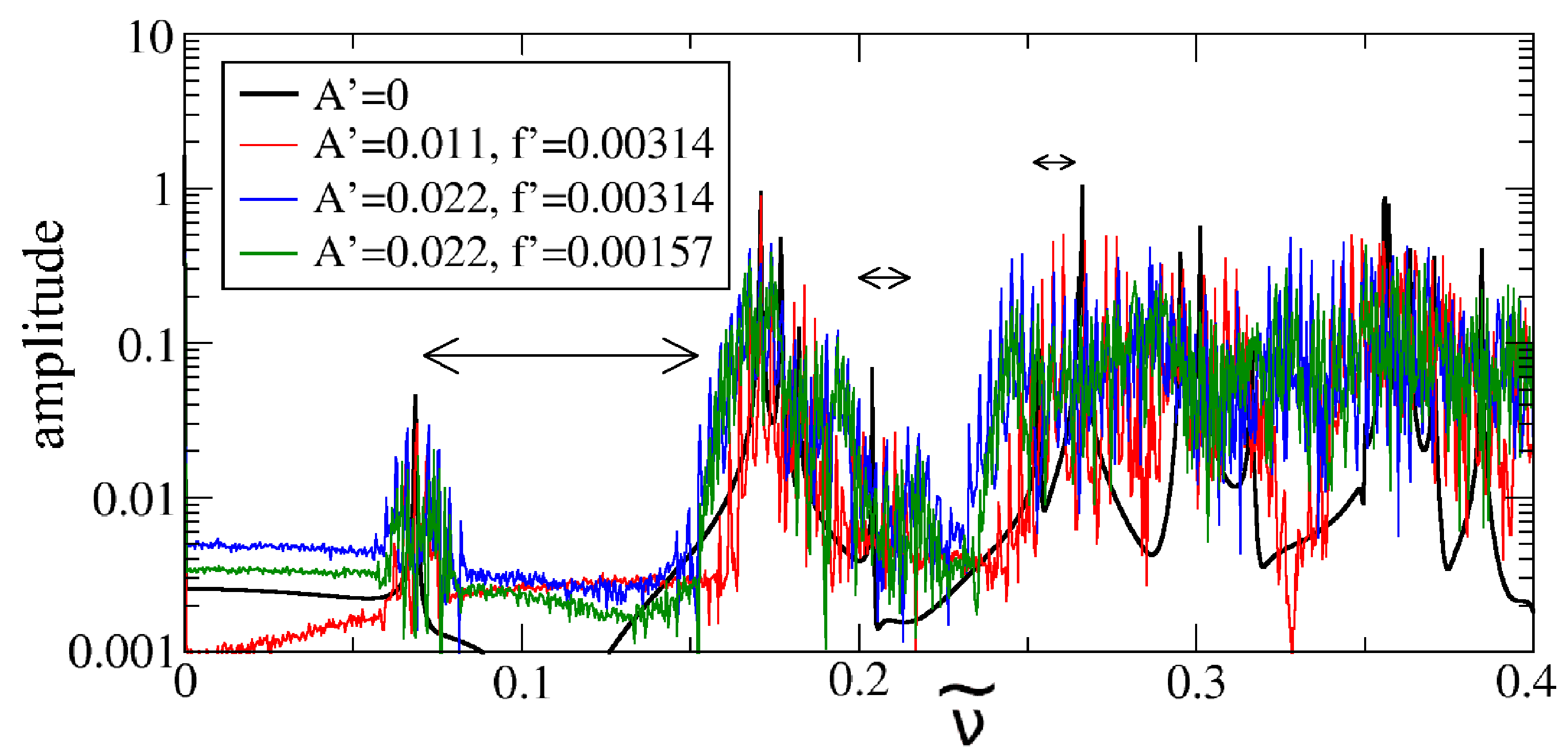}
        \caption{Frequency spectrum at $k=0$ for a point-source excitation for a system with, an Fe hollow
	cylinder
	in an epoxy matrix with dimensions as shown in the outlines of Fig.~\ref{A5}. Shown is the spectrum without
        any time variation of cylinder radius ($A^\prime=0$)
        as well as with nonzero maximum amplitude modulations $A^\prime$, at two different frequencies
        (dimensionless units). The horizontal arrows show the locations of the absolute
	band gaps in the non-time-varying case (\textit{viz.} Fig.~\ref{A3}).}
\label{A4}
\end{figure}

As we time-vary the outer radius of the cylinder, we obtain the spectrum shown in
Fig.~\ref{A4} for different maximum amplitude variations and modulation frequencies, scaled
respectively to be dimensionless as $A^\prime=A/a$ and $f^\prime=fa/c,$ at
$k=0$. The vertical axis depicts the amplitude of the frequency spectrum
$\sqrt{u_x(\tilde{\nu})^2+u_y(\tilde{\nu})^2},$ obtained after Fourier-transforming
the value of the displacement at a fixed point inside the cylinder, near its centre. The
horizontal arrows denote
the location of the absolute band gaps in the non-time-varying case (\textit{viz.} Fig.~\ref{A3}).

In Fig.~\ref{A5} we show the displacement profiles of the system for the points Z1-Z4 at $k=0$
in Fig.~\ref{A3}. The
point Z2 actually corresponds to four nearly-degenerate frequencies. Specifically, we display
a plane projection of the displacement $u_x^2+u_y^2$ of a point inside the
ring in order to visualize the localization of the vibrations.

The results are very similar to the corresponding case of a solid cylinder. The displacement
profile of point Z1 has the same character as that of point X1 (\textit{viz.} Fig.~\ref{localFeEp}a).
Point Z4 here corresponds to the displacement profile of point X2 in that case (\textit{viz.}
Fig.~\ref{localFeEp597}a). Point Z2 has a strong lifting of its degeneracy and only two of the four
closely-spaced modes show the same localization
as in the solid cylinder case: Z2(2) and Z2(3) which correspond to the two
nearly-degenerate modes located at the point Y2 in Fig.~\ref{localSiEp}d for the Si
cylinder in epoxy. In Fig.~\ref{localSiEp}d we
showed only one of these (corresponding to Z2(2) here). The
profiles Z2(1) and Z2(4) show symmetries not seen in the case of the solid cylinder
showing strong indications of mixing with the higher modes Z3 and Z4.

\end{document}